\documentclass{appolb}
\usepackage{epsfig}
\def\GeV{{\rm GeV}}
\begin{document}
%\date{\today}
\pagestyle{plain}
%% uncomment the following line to get equations numbered by (sec.num)
%\eqsec
\newcount\eLiNe\eLiNe=\inputlineno\advance\eLiNe by -1
\title{Update of MRST Parton Distributions.}
\author{R.S. Thorne
\address{Cavendish Laboratory, University of 
Cambridge\footnote{Royal Society University Research Fellow}, 
Madingley Road, \\
Cambridge, CB3 0HE, U.K.}\\
\bigskip
A.D. Martin and W.J. Stirling
\address{Department of Physics, University of 
Durham, Durham, DH1 3LE, U.K.}\\
\bigskip
R.G. Roberts
\address{Theory Division, CERN, 1211 Geneva 23, Switzerland}}
\maketitle

\begin{abstract}
We discuss the latest update of the MRST parton distributions in response 
to the most recent data. We discuss the areas where there are hints of 
difficulties in the global fit, and compare to some other updated 
sets of parton distributions, particularly CTEQ6. 
We briefly discuss the issue of uncertainties associated with partons.   
\end{abstract}

Over the past couple of years there has been a large amount of updated 
data both from HERA \cite{H1,ZEUS}, on small $x$ structure functions, 
and from the Tevatron,
on high-$E_T$ jet production \cite{D0,CDF}, 
which has been more accurate than previous 
data, and expanded the phase space significantly. This had led to a number of
updated sets of parton distributions \cite{MRST01}-\cite{zeusfit}. 
In this talk we discuss the 
most recent updates to the MRST set of parton distributions, highlighting the
successes and failings, and also compare to other new sets of  
distributions.  

The updated MRST partons were released in 2001 \cite{MRST01}. Compared to
previous sets the main improvement was in the accuracy of the determination 
of the gluon distribution, which was constrained far more strongly at high 
$x$ due to the new Tevatron jet data. $\alpha_S(M_Z^2)$ was left as a free
parameter in the fit, and found to be $0.119 \pm 0.002({\rm exp})\pm 0.003
({\rm theory})$,
where the experimental error was determined by letting the $\chi^2$ for the 
global fit increase by 20 from the minimum
(see \cite{delchi} for a discussion of the suitable 
increment in $\chi^2$ to determine the error in a global fit). 
The fit was of good quality overall, but 
struggled a little in some regions. It was hard to provide enough high $x$ 
gluon to fit the jet data very well, and also to have sufficient 
moderate $x$ gluon 
to obtain a large enough value of $dF_2(x,Q^2)/d\ln Q^2$ for $x \sim 0.01$. 
Conversely, the data required the very small $x$ gluon to be small (which
also helps the previous shortcomings due to the constraint on the total 
gluon from the momentum sum rule), and at our input scale $Q_0^2=1\GeV^2$
it was found to be necessary to expand our parameterization to allow the very 
small $x$ gluon to become negative. This latter point led to a dangerously 
small prediction for $F_L(x,Q^2)$ at small $x$ and $Q^2$.     

\vspace{-0.6cm}
\begin{figure}[htbp]
\includegraphics[width=6cm]{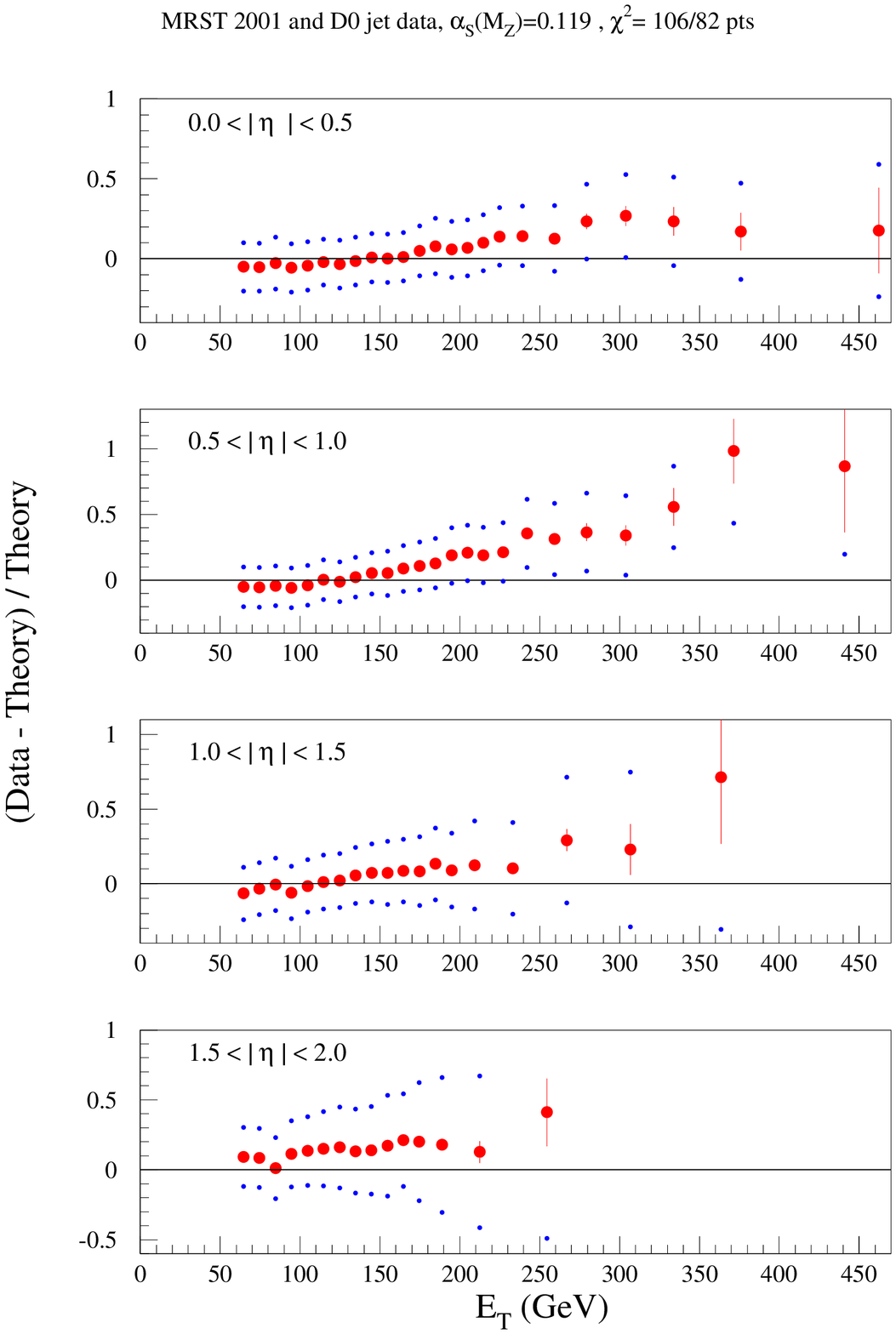}
\hspace{0.5cm}
\includegraphics[width=6cm]{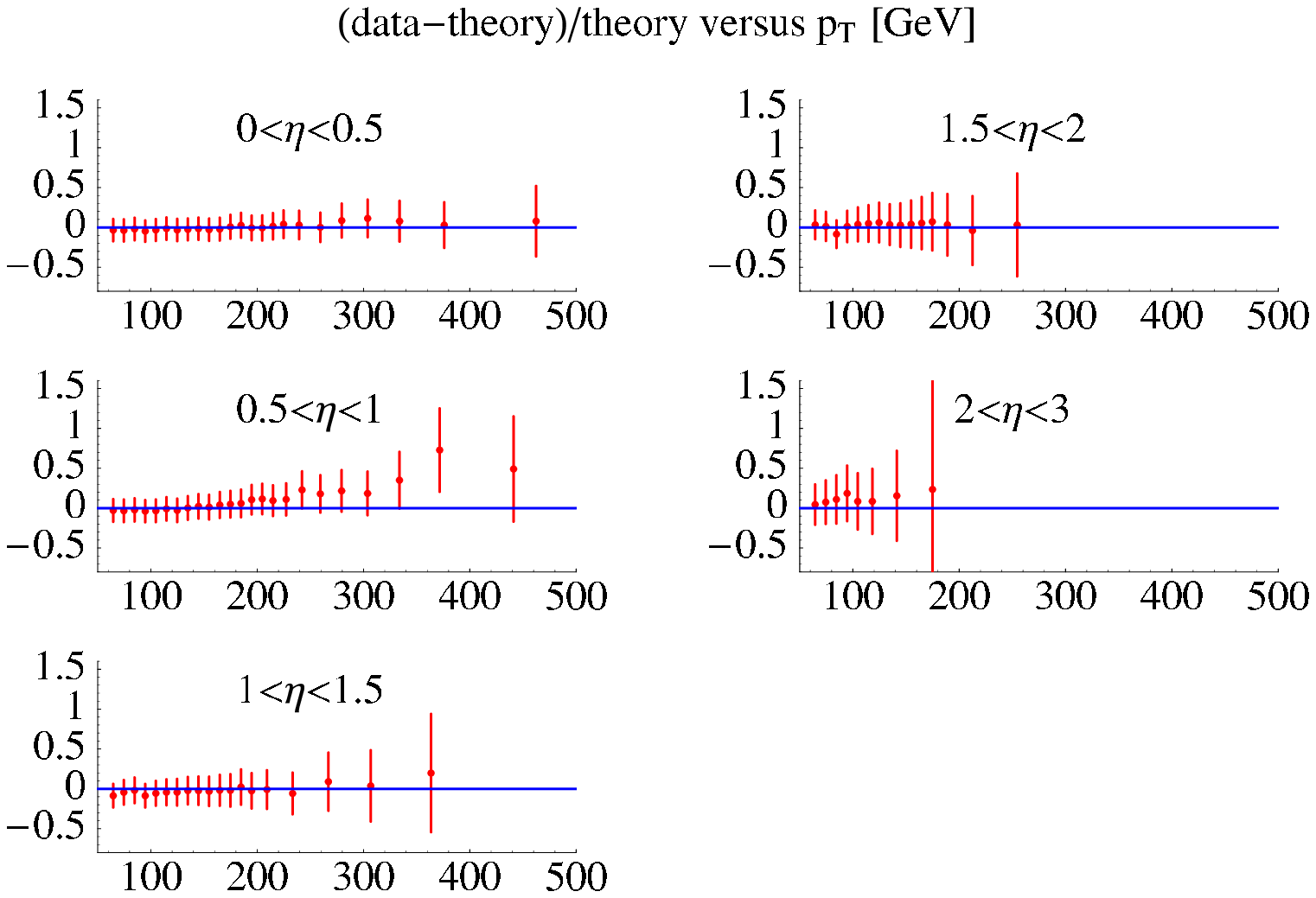}
\vspace{-0.6cm}
\caption{Comparison of MRST fit to D0 jet data to CTEQ6 fit.}
\label{jets}
\end{figure}

Soon afterwards the CTEQ6 set of partons was published \cite{CTEQ6}. In most 
ways these are very similar to the MRST01 partons, and produce similar 
results. However, there are a number of significant differences, 
particularly concerning the gluon. CTEQ have developed a different type of 
parameterization for the partons, which allows for a different shape at
very high $x$. Whereas MRST were only able to get a completely 
satisfactory fit to the Tevatron jet data if the input gluon is allowed 
to have a definite kink 
at $x\sim 0.5$ (and with $\alpha_S(M_Z^2)=0.121$), CTEQ 
obtain a very good fit with no such modifications.    

However, this problem of obtaining a very good fit to the jet data depends 
on many issues. CTEQ do indeed obtain a much better fit 
using this new parameterization for the gluon   
(with same NLO prescription the $\chi^2$ quality is about 50 better)
as seen for D0 data in Fig. 1. 
However, there are many differences in their 
approach compared to MRST other than the parameterization: 
CTEQ cut data above $Q^2 = 4 \GeV^2$, compared to 
$Q^2 = 2 \GeV^2$; they do not use some data sets used in \cite{MRST01}, 
i.e. SLAC and one H1 high-$Q^2$ set;
they use ($10\%$) systematic errors (in quadrature) 
for Drell-Yan data whereas in \cite{MRST01} only statistical errors are used.
Additionally CTEQ have a positive-definite small $x$ gluon at their starting 
scale of $Q_0^2= 1.69 \GeV^2$, they use 
a massless charm prescription and there are various other minor differences.

In order to investigate which initial choices are most important for
the quality of the fit to the jet data, or equivalently, which affect the 
extracted form of the high-$x$ gluon, we performed various fits changing
these choices.  
We found that we can improve the fits to jets within the global fit by 
various modifications. Unexpectedly, 
allowing  one of the parameters controlling the negative contribution to our 
gluon at very small $x$ to vary away from a previously fixed value resulted in 
$\Delta \chi^2_J \sim -5$.  
The fit to the Drell-Yan data actually competes with that to the jets, 
and using only statistical errors (the systematic errors being defined a 
little vaguely) presumably overweights these. Adding $5\%$ systematic 
errors in quadrature to the statistical errors (which is probably the 
best approach \cite{DrellYan}) leads to $\Delta \chi^2_J \sim -10$.
Both these modifications should be performed, and will be implemented in 
future MRST fits. The resulting partons are currently denoted MRST$\star$. 
The only real change compared to MRST01 is for the high $x$ gluon. 

\begin{figure}[htbp]
\begin{center}
\includegraphics[width=7cm]{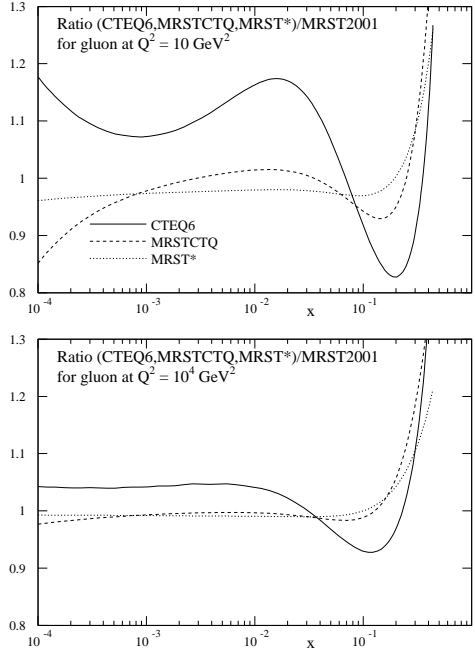}
\end{center}
\vspace{-1.8cm}
\caption{Comparison of MRST2001 gluon distribution to the other 
distributions described in the text.}
\label{ratg}
\end{figure}

We also discovered that further changes could improve the quality of the 
jet fit. Changing the $Q^2$-cut on the data from the MRST value of  
$Q^2 = 2 \GeV^2$ to the CTEQ value of $Q^2 = 4 \GeV^2$ 
leads to $\Delta \chi^2_J \sim -10$.     
Fitting to the same data as CTEQ, i.e. omitting the SLAC data and one
H1 high-$Q^2$ data set and increasing the Drell-Yan systematic errors to 
$10\%$ leads to  $\Delta \chi^2_J \sim -15$.
The cumulative effect of all these 
above steps in a single fit is $\Delta \chi^2_J 
\sim -40$, which is obtained with a smooth high-$x$ gluon. 
We denote the resulting partons by MRSTCTQ. We conclude that 
the remaining improvement of $\Delta \chi^2_J \sim -10$ seen by CTEQ is due 
mainly to their new parameterization, but that this is only a relatively 
minor effect. Indeed, we compare the gluons from CTEQ6, MRST01, MRST$\star$ 
and MRSTCTQ in Fig. 2. Clearly MRSTCTQ has a very similar high-$x$ gluon to 
CTEQ6, and even MRST$\star$ has gone much of the way in the same direction. 
However, all the MRST gluons are different 
from the CTEQ6 gluon at smaller $x$ due to their freedom to have a 
negative input distribution. We also note that although we feel the steps 
producing the MRST$\star$ partons should be made in future, the 
further ones leading to MRSTCTQ are a different matter. 
Although they improve the 
quality of the jet fit they are not the best fit when including the 
data sets omitted 
by CTEQ and the fit is not good at all for data with $Q^2 < 4 \GeV^2$.
It is certainly true that we should question the nature of our cut on $Q^2$
(as well as on $W^2$ and $x$), but this is a complicated question which will 
be addressed elsewhere \cite{cuts}. 
 
The comparison to the other sets of parton distributions obtained by fits to 
mainly structure function data, e.g. \cite{h1fit}-\cite{zeusfit}, are 
qualitatively the same as they have been for some time. Since only 
MRST and CTEQ fit the jet data, it is only these partons which have a 
direct constraint on the high-$x$ gluon. All other fits always obtain, to 
varying degrees, a smaller high-$x$ gluon which consequently 
allows both a larger
moderate-$x$ gluon to fit the HERA data and a usually a slightly 
smaller value of $\alpha_S(M_Z^2)$. Hence, the omission of the jet data 
tends to mask slightly the possible problems encountered in trying to fit the 
HERA data very well. 

Recently, many groups have not only obtained partons from a best 
fit but, using various methods, have also examined the uncertainty on 
these partons due to experimental errors. MRST have 
concentrated on the Lagrange Multiplier technique \cite{lmethod} in order to
obtain uncertainties on physical quantities and the corresponding extreme 
sets of partons. Such uncertainties, and partons, are available for $W$ and
Higgs production at the Tevatron and LHC, and for charged current 
cross-sections at $x=0.5$ for HERA \cite{lagrange}, both for fixed and 
varying $\alpha_S$. However, we have always 
believed that theory is one of the dominant sources of error. Hence, 
as well as attempting to determine 
the areas where the current theory may require corrections 
by investigating the cuts on data \cite{cuts}, we have also 
produced approximate NNLO parton distributions and predictions \cite{MRSTNNLO}
(based on the approximate splitting functions
\cite{NNLOsplit} obtained from the known NNLO moments \cite{NNLOmoms}).
Indeed, we find, for example,  
that the NNLO $W$ cross-section at the Tevatron is $4\%$ higher than at 
NLO, and believe this result is reliable.
This change is at least as large as the uncertainty due to experimental 
errors, and $W$ production  
is likely to be subject to smaller theoretical uncertainty
than many other quantities - particularly those directly related to the 
gluon. Hence, 
an understanding of theoretical uncertainties seems to be a priority 
at present.

\end{document}